\documentclass[twocolumn,10pt,aip,graphicx]{revtex4-1}
\usepackage{graphicx}
\usepackage{dcolumn}
\usepackage{bm}
\usepackage[utf8]{inputenc}
\usepackage[T1]{fontenc}
\usepackage{mathptmx}
\usepackage{etoolbox}
\usepackage{subfigure}
\usepackage{amsmath}
\usepackage{xcolor}

\makeatletter
\def\@email#1#2{%
 \endgroup
 \patchcmd{\titleblock@produce}
  {\frontmatter@RRAPformat}
  {\frontmatter@RRAPformat{\produce@RRAP{*#1\href{mailto:#2}{#2}}}\frontmatter@RRAPformat}
  {}{}
}%
\makeatother
\begin{document}

\title[]{Indirect nonlinear interaction between toroidal Alfv\'en eigenmode and ion temperature gradient mode mediated by zonal structures}
\author{Qian Fang}
\affiliation{Institute for Fusion Theory and Simulation and Department of Physics, Zhejiang University, Hangzhou 310027, China}

\author{Guangyu Wei}
\affiliation{Institute for Fusion Theory and Simulation and Department of Physics, Zhejiang University, Hangzhou 310027, China}

\author{Ningfei Chen}
\affiliation{Institute for Fusion Theory and Simulation and Department of Physics, Zhejiang University, Hangzhou 310027, China}

\author{Liu Chen}
\affiliation{Institute for Fusion Theory and Simulation and Department of Physics, Zhejiang University, Hangzhou 310027, China}
\affiliation{Center for Nonlinear Plasma Science and C.R. ENEA Frascati, Via E. Fermi 45, 00044 Frascati, Italy}
\author{Fulvio Zonca}
\affiliation{Center for Nonlinear Plasma Science and C.R. ENEA Frascati, Via E. Fermi 45, 00044 Frascati, Italy}
\affiliation{Institute for Fusion Theory and Simulation and Department of Physics, Zhejiang University, Hangzhou 310027, China}
\author{Zhiyong Qiu}\email{zqiu@ipp.ac.cn}
\affiliation{Key Laboratory of Frontier Physics in Controlled Nuclear Fusion and Institute of Plasma Physics, Chinese Academy of Sciences, Hefei 230031, China}
\affiliation{Center for Nonlinear Plasma Science and C.R. ENEA Frascati, Via E. Fermi 45, 00044 Frascati, Italy}

\date{\today}

\begin{abstract}
The indirect nonlinear interactions between toroidal Alfv\'en eigenmode (TAE) and ion temperature gradient mode (ITG) are investigated using nonlinear gyrokinetic theory and ballooning mode formalism. More specifically, the local nonlinear ITG mode equation is derived adopting the fluid-ion approximation, with the contributions of zonal field structure and phase space zonal structure beat-driven by finite amplitude TAE accounted for on the same footing. The obtained nonlinear ITG mode equation is solved both analytically and numerically, and it is found that, the zonal structure beat-driven by TAE has only weakly destabilizing effects on ITG, contrary to usual speculations and existing numerical results. 
\end{abstract}

\maketitle

\section{Introduction}

Drift wave (DW) turbulence \cite{WHortonRMP1999} and shear Alfv\'en wave (SAW) \cite{YKolesnichenkoVAE1967, AMikhailovskiiSPJ1975, MRosenbluthPRL1975, LChenPoP1994, AFasoliNF2007, LChenRMP2016} are two major categories of collective oscillations contributing to anomalous cross-field transport in magnetized plasmas. DWs are typically micro-scale electrostatic oscillations driven by free energy associated with thermal plasma density and/or temperature nonuniformities, and are widely accepted as candidates for inducing bulk plasma transport. On the other hand, SAW instabilities or, more precisely, Alfv\'en eigenmodes (AEs) due to equilibrium magnetic geometry excited by energetic particles (EPs), are expected to play crucial role in EP transport. DW and SAW instabilities are characterized by different spatiotemporal scales, driven unstable by different free energy sources and dominate transport of different energy regime, and, thus, are typically studied separately. However, it is expected that there are complex cross-scale couplings among DWs and SAWs in burning plasmas, due to the mediation by EPs \cite{FZoncaPPCF2015, LChenRMP2016}. Specifically, on the one hand, EPs drive meso-scale SAW instabilities that can provide nonlinear feedback to both the macroscopic plasma profiles and microscopic fluctuations. On the other hand, EPs can linearly and nonlinearly (via SAWs) excite zonal structures (ZS) \cite{RNazikianPRL2008, ZQiuNF2016, ZQiuNF2017}, thus act as generators of nonlinear equilibrium \cite{MFalessiNJP2023, FZoncaJPCS2021}. There are now raising interest in their mutual effects on each other, due to the recent observed thermal plasma confinement improvement in the presence of EPs \cite{DSienaPRL2021, HHanNature2022}. E.g., experiments and simulations have indicated that the presence of EPs can significantly suppress ion temperature gradient (ITG) turbulence, thereby reducing ion thermal stiffness and enhancing ion confinement \cite{HHanNature2022}. Early linear gyrokinetic simulations suggested that the presence of EPs might trigger internal transport barriers \cite{MRomanelliPPCF2010} (ITBs) formation, which could decrease core turbulence levels and improve bulk plasma confinement.

There are several channels for linear stabilization of micro-turbulence by EPs via, dilution of destabilizing bulk ions \cite{GTardiniNF2007}, EP-pressure induced Shafranov shift stabilization \cite{CBourdelleNF2005}, resonant stabilization \cite{DSienaNF2018, NBonanomiNF2018}, electromagnetic stabilization \cite{MRomanelliPPCF2010, JCitrinPPCF2014}, and so on. The linear stabilization effects are now relatively well understood, and interested readers may refer to a recent review for a more complete picture\cite{JCitrinPPCF2023}.

Moreover, numerical studies have revealed a nonlinear mechanism for fast-ion-enhanced EM-stabilization of ITG \cite{JCitrinPRL2013}, which could be much more stabilizing than linear ones \cite{NBonanomiNF2018}, then these nonlinear coupling mechanisms have drawn significant research interest \cite{VMarchenkoPoP2022}. A potential physical interpretation is that the presence of EPs provides marginally linearly stable EM-modes, i.e., toroidal Alfv\'en eigenmodes (TAEs) \cite{CChengAP1985} that transfer energy nonlinearly to ZS \cite{AHasegawaPoF1979, MRosenbluthPRL1998}, and the increase in ZS levels stabilizes the ion-scale turbulence \cite{VMarchenkoPoP2022, SMazziNP2022}. However, the underlying physical processes of this issue remain to be explored.

More recently, the direct nonlinear interactions among TAE and electron drift wave (eDW) are proposed and analyzed. It is found that direct nonlinear scattering by ambient eDWs may significantly reduce or even completely suppress TAE stability, due to radiative  damping of the nonlinearly generated high-$n$ KAW quasi-modes \cite{LChenNF2022}. On the other hand, for typical reactor  parameters and fluctuation intensity, the ``inverse'' process, i.e., the direct nonlinear scatterings of eDW by ambient TAEs, have negligible net effects on the eDW stability \cite{LChenNF2023}, which differs from the results of previous simulations. To understand the bulk plasma confinement in the presence of EPs, as a natural step forward, in this work, we will investigate the in-direct nonlinear interaction among TAE and DWs mediated by ZS, i.e., the ``linear'' stability of DWs in the presence of ZS nonlinearly excited by TAE \cite{LChenPRL2012}. 

ZS, including zonal flows (ZF), zonal currents (ZC), and more generally, phase space zonal structures (PSZS) \cite{FZoncaNJP2015,MFalessiNJP2023}, are known to play important roles in regulating micro-scale DW type instabilities including drift Alfv\'en waves (DAWs) \cite{LChenPoP2000, ZLinScience1998, LChenPRL2012}. The regulation is achieved via the nonlinear generation of ZS by DWs/DAWs, and in the process, ZS scatters DWs/DAWs into linearly stable short radial wavelength domain. The ZS excitation can be achieved via the spontaneous excitation via radial envelope modulation of DWs/DAWs \cite{LChenPoP2000, LChenPRL2012}, as well as beat-driven process of DW/DAW self-coupling frequently observed in large scale simulations, characterized by a ZF growth rate being twice of DW/DAW instantaneous growth rate \cite{YTodoNF2010, ABiancalaniJPP2020}.  
To simplify the analysis while focusing on the main physics picture of nonlinearly generated ZS on DW stability, in this work, we consider only the beat-driven process by TAE \cite{ZQiuPoP2016, LiuChenWLS2023}, without including the spontaneous excitation process. It is already shown analytically and numerically that, zero frequency radial electric field can significantly stabilize ITG via modification of the ``potential well'' depth \cite{NChenPoP2021}. 
Here, using ITG as the paradigmatic model and following the analysis of Ref. \citenum{NChenPoP2021}, we will investigate the indirect nonlinear interaction between TAE and DW, mediated by ZS. Technically, this is achieved by deriving the expression of TAE beat-driven ZS, and taking it as a ``nonlinear'' equilibrium into the ITG eigenmode equation. Thereafter, the local dispersion relation and mode structure of the ITG under the influence of ZS are derived in the ballooning space. Here, ``local'' means the ITG eigenmode equation is solved along the magnetic field lines, while physics associated with radial envelope modulation are neglected systematically.

The rest of the paper is organized as follows. In section \ref{Theoretical model}, the theoretical model and governing equations are presented. In section \ref{Nonlinear ZFZS generation}, we analyzed the nonlinear generation of ZS in detail. In section \ref{Effects of beat-driven ZFZS on DW linear stability}, we incorporated the description of beat-driven ZS into the existing linear dispersion relation to investigate the ``nonlinear'' effects on ITG linear stability. In section \ref{Analytical and numerical results}, analytical and numerical results are discussed in both the short- and long-wavelength limits. Finally, a brief summary and discussions are given in section \ref{Summary and discussion}.

\section{Theoretical model}\label{Theoretical model}

For simplicity and clarity of discussion, we consider a low-$\beta$ tokamak with large aspect ratio and concentric circular magnetic surfaces, in which the equilibrium magnetic field is given by $\mathbf{B}=B_{0}\left[\mathbf{e}_{\zeta}/\left(1+\epsilon\cos\theta\right)+\epsilon\mathbf{e}_{\theta}/q\right]$.
Here, $\epsilon\equiv r/R<1$ is the inverse aspect ratio, $r$ and $R$ are, respectively, the minor and major radii of the torus, $\beta\equiv 8\pi P/B_{0}^{2}\sim\mathcal{O}\left(\epsilon^{2}\right)\ll1$
is the ratio between plasma and magnetic pressure, $\zeta$ and $\theta$ are the toroidal and poloidal angles, respectively, and $q\equiv rB_{\zeta}/(RB_{\theta})$ is the safety factor. 
To simplify the analysis while focuing on main physics, the equilibrium distribution functions of thermal plasmas are taken as local Maxwellian, while trapped particle effects are systematically neglected.
We take $\delta\phi$ and $\delta A_{\parallel}$ as the field variables in the low-$\beta$ limit, where $\delta\phi$ is the scalar potential and $\delta A_{\parallel}$ is the parallel component of vector potential, i.e., $\delta\mathbf{A}\simeq\delta A_{\parallel}\mathbf{b}$ and $\mathbf{b}\equiv\mathbf{B}/B$ is the unit vector along equilibrium magnetic field. 
An alternative field variable $\delta\psi\equiv\omega\delta A_{\parallel}/(ck_{\parallel})$ is also adopted for $n\neq0$ TAEs, and one has $\delta\psi=\delta\phi$ in the ideal MHD limit.

In gyrokinetic theory, the perturbed distribution function, $\delta f_{s}$, with the subscript $s$ representing the particle species $s=e,\ i$, can be denoted as
\begin{eqnarray}
\delta f_{s} & = & -\left(\frac{e}{T}\right)_{s}\delta\phi F_{Ms}+\exp\left(-\mathbf{\rho}_{s}\cdot\nabla\right)\delta H_{s},\label{perturbed distribution function}
\end{eqnarray}
and its non-adiabatic component $\delta H_s$ obeys the nonlinear gyrokinetic equation \cite{EFriemanPoF1982}
\begin{equation}
\begin{aligned}\left(\partial_{t}+v_{\parallel}\partial_{l}+i\omega_{Ds}\right)\delta H_{s}= & -i\left(\frac{e}{T}\right)_{s}\left(\omega-\omega_{*}^{t}\right)F_{Ms}J_{k}\delta L_{k}\\
 & -\sum_{\mathbf{k}=\mathbf{k^{'}+\mathbf{k^{''}}}}\Lambda_{k^{''}}^{k^{'}}J_{k^{'}}\delta L_{k^{'}}\delta H_{k^{''}}.
\end{aligned}
\label{nonlinear gyrokinetic equation}
\end{equation}
Here, $F_{Ms}$ is the Maxwellian distribution representing equilibrium thermal particle distribution, $\mathbf{\rho}_{s}=\mathbf{b}\times\mathbf{v}/\omega_{cs}$ is the Larmor radius with $\omega_{cs}\equiv eB_{0}/\left(m_{s}c\right)$
being  the cyclotron frequency, $l$ is the arc length along the equilibrium magnetic field line,   $\omega_{Ds}=2\hat{\omega}_{ds}C$ represents the magnetic drift frequency, with $\hat{\omega}_{ds}\equiv\omega_{ds}(x_{\perp}^{2}/{2}+x_{\parallel}^{2})$, $\omega_{ds}\equiv k_{\theta}cT_{s}/(e_{s}B_{0}R)$, $x_{\perp}=v_{\perp}/v_{ts}$ and $x_{\parallel}=v_{\parallel}/v_{ts}$ being the particle perpendicular/parallel velocities normalized by thermal velocity $v_{ts}\equiv\sqrt{2T_{s}/m_{s}}$, respectively,  and $C\equiv\cos\theta-\sin\theta k_{r}/k_{\theta}$ is related to the magnetic  curvature, with $k_r$ and $k_{\theta}=-m_0/r_0$ being the radial/poloidal wavenumbers.  Furthermore, $\omega_{*s}^{t}\equiv\omega_{*s}(1+\eta_{s}(v^{2}/v_{ts}^{2}-3/2))$ is the diamagnetic drift frequency associated with plasma nonuniformity, with $\omega_{*s}=k_{\theta}cT_{s}/\left(e_{s}B_0L_{n}\right)$, $\text{\ensuremath{\eta_{s}=L_{ns}/L_{Ts}}}$,
where $L_{n}\equiv-N/\left(\partial N/\partial r\right)$, $L_{T}\equiv-T/\left(\partial T/\partial r\right)$ are
the characteristic lengths of density and temperature nonuniformities, respectively,  $J_{k}=J_{0}\left(k_{\perp}\rho_{s}\right)$ is the Bessel function of zero index accounting for finite Larmor radius (FLR) effects and $k_{\perp}=\sqrt{k_{r}^{2}+k_{\theta}^{2}}$ is perpendicular wavenumber. Furthermore, $\Lambda_{k^{''}}^{k^{'}}\equiv(c/B_{0})\mathbf{b}\cdot(\mathbf{k^{''}}\times\mathbf{k^{'}})$, with  $\sum_{\mathbf{k}=\mathbf{k^{'}+\mathbf{k^{''}}}}\Lambda_{k^{''}}^{k^{'}}$ representing the selection rule of frequency and wavenumber matching conditions for nonlinear mode coupling, $\delta L_{k}\equiv\delta\phi_{k}-k_{\parallel}v_{\parallel}\delta\psi_{k}/\omega_{k}$ is the scalar potential in guiding-center moving frame, and other
notations are standard. 
In the following analysis, the non-adiabatic particle response $\delta H_{s}$ can be separated into linear and nonlinear components, i.e., $\delta H\equiv\delta H^{L}+\delta H^{NL}$, with the superscripts ``L'' and ``NL'' denoting the linear and nonlinear components, respectively, and can be derived perturbatively assuming $\delta H^{NL}\ll\delta H^{L}$. 

In the $\beta\ll 1$ limit with negligible magnetic compression, the nonlinear gyrokinetic equation can be closed by charge quasi-neutrality 
\begin{eqnarray}
\frac{Ne^{2}}{T_{i}}\left(1+\frac{1}{\tau}\right)\delta\phi_{k} & = & \sum_s\left\langle e_{s}J_{k}\delta H_{k}\right\rangle,\label{quasineutrality condition}
\end{eqnarray}
and parallel Ampere's law 
\begin{equation}
-\frac{c}{4\pi}\nabla_{\perp}^{2}\delta A_{\parallel}=\delta J_{\parallel}=\left\langle -ev_{\parallel}\delta f_e\right\rangle.\label{parallel Ampere's law}
\end{equation}
Here, terms on the left hand side of equation (\ref{quasineutrality condition}) are contributions from adiabatic responses of ion and electron, respectively, $\tau\equiv T_{e}/T_{i}$ is the electron to ion temperature ratio, and $\langle \cdot\cdot\cdot\rangle$ represents velocity space integration.

We now consider the effects of indirect modulation of ITG stability by finite amplitude TAE, mediated by beat-driven ZS. More specifically, this is a two-step process, with the first process being finite amplitude TAE $\Omega_{0}(\ensuremath{\omega_{0}},\ensuremath{k_{\theta0}})$
couples with its complex conjugate $\Omega_{0}^{*}(\ensuremath{-\omega_{0}},\ensuremath{-k_{\theta0}})$
to generate ZS $\Omega_{Z}(\ensuremath{\omega_{Z}},\ensuremath{k_{r Z}})$
with $\omega_{Z}\approx0$, $k_{\parallel Z}=0$, $k_{\theta Z}=0$,
and finite $k_{r Z}$, which is analysed in section \ref{Nonlinear ZFZS generation}. In the second process, the beat-driven ZS affects the linear stability of ITG
$\Omega_{I}(\ensuremath{\omega_{I}},\ensuremath{k_{\theta I}})$, and is analyzed in section \ref{Effects of beat-driven ZFZS on DW linear stability}.
Here, the subscripts $0$, $\ensuremath{Z}$ and $I$ denote quantities
associated with the TAE, ZS and ITG, respectively. For DWs, especially ITG, with most unstable modes characterized with high mode numbers, and the characteristic scale
of equilibrium profile variation generally much larger than the distance
between neighbouring mode rational surfaces, the well-known ballooning-mode
decomposition \cite{JConnorPRL1978} can be adopted to decompose the  two-dimensional  eigenvalue problem into two coupled one-dimensional eigenvalue problems, i.e., 
\begin{eqnarray}
\delta\phi_I & = & A_{I}e^{-in\zeta-i\omega t}\underset{j}{\sum}\Phi_I\left(x-j\right)e^{i(m_0+j)\theta}+c.c.\label{ballooning-mode}
\end{eqnarray}
Here, $A_I$ is the radial envelope of ITG, $n$
is the toroidal mode number, $(m=m_{0}+j$) is the poloidal mode number
with $m_{0}$ being its reference value satisfying $nq(r_{0})=m_{0}$, and $r_{0}$ denotes the plasma radial position about which the ITG is assumed to be localized.
Furthermore, $x=nq-m_0\simeq nq^{'}(r-r_0)$, $\left|j\right| \ll m_0$ is an integer, and  $\Phi_{I}\left(x-j\right)$ is the fine radial structure due to finite $k_{\parallel}$, with $\int\left|\Phi_{I}\right|^2dx=1$ being normalization condition. In the present analysis, we however, will focus on the local stability of ITG, as the ZS driven by TAE are expected to have a larger radial scale than ITG \cite{ZQiuNF2017}. The global problem with the radial envelope modulation will be analyzed in a future publication.

\section{Nonlinear ZS generation}\label{Nonlinear ZFZS generation}

In this section, based on the gyrokinetic theoretic framework, i.e., equations (\ref{nonlinear gyrokinetic equation})-(\ref{parallel Ampere's law}),
the nonlinear generation of ZS by TAEs self-beating is investigated.
Briefly speaking, the particle responses to ZS due to TAE self-coupling (i.e., PSZS) are derived from the nonlinear gyrokinetic equation, which are then substituted into the quasi-neutrality condition and parallel Ampere's law to derive the zonal scalar and vector potential response, which then will be used to study the TAE effects on ITG via ZS mediation in section \ref{Effects of beat-driven ZFZS on DW linear stability}.

We start from the linear particle response to the finite amplitude TAE $\Omega_{0}(\ensuremath{\omega_{0}},\ensuremath{k_{\theta 0}})$,
which will be used in the nonlinear term of the gyrokinetic equation
later. For massless-electron with $\left|k_{\parallel}v_{te}/\omega_{0}\right|\gg1$, $\left|k_{\perp}\rho_{e}\right|\ll1$ and $J_0\left(k_{\perp}\rho_{e}\right)\sim1$
satisfied, the linear electron response to $\Omega_{0}$ can be straightforwardly
derived as
\begin{eqnarray}
\delta H_{0,e}^{L} & \simeq & -\frac{e}{T_{e}}\left(1-\frac{\omega_{*e,0}^{t}}{\omega_{0}}\right)F_{Me}\delta\psi_{0}.\label{electron linear response of 0}
\end{eqnarray}
Meanwhile, for ions with $\left|k_{\parallel}v_{ti}/\omega_{0}\right|\ll1$, we have
\begin{eqnarray}
\delta H_{0,i}^{L} & \simeq & \frac{e}{T_{i}}\left(1-\frac{\omega_{*i,0}^{t}}{\omega_{0}}\right)F_{Mi}J_{0}\delta\phi_{0}.\label{ion linear response of 0}
\end{eqnarray}
Substituting above expressions into the quasi-neutrality condition and vorticity equation \cite{LChenRMP2016}, the dispersion relation of TAE can be obtained. In our analysis focusing on TAE effects on ITG stability, we however, do not need the detailed expression of TAE dispersion relation.

The linear and nonlinear particle responses to ZS can be derived by transforming equation (\ref{nonlinear gyrokinetic equation}) into drift orbit center coordinate, i.e., taking $\delta H_{Z,s}\equiv\delta H_{dZ,s}\exp\left(i\lambda_{Z}\right)$, with $\lambda_{Z}\equiv\hat{\lambda}_{dZ}\cos\theta$ satisfying $\omega_{tr}\partial_{\theta}\lambda_{Z}+\omega_{Drs}=0$, and $\hat{\lambda}_{dZ}\equiv-2\hat{\omega}_{dr}/\omega_{tr}$ being the normalized drift orbit width.  Here, $\omega_{tr}\equiv v_{\parallel}/qR$ is the transit frequency, $\omega_{Drs}\equiv-2\hat{\omega}_{drs}\sin\theta$, and $\hat{\omega}_{drs}\equiv k_{r}cT_{s}(x_{\perp}^{2}/2+x_{\parallel}^{2})/(e_{s}B_{0}R)$. Substituting $\delta H_{Z,s}$
into equation (\ref{nonlinear gyrokinetic equation}), we have
\begin{eqnarray}
\left(\partial_{t}+\omega_{tr}\partial_{\theta}\right)\delta H_{dZ}&=&-i\left(\frac{e}{T}\right)_{s}e^{-i\lambda_{Z}}\omega_{Z}F_{Ms}J_{Z}\left(\delta\phi-\frac{v_{\parallel}}{c}\delta A\right)_{Z}\nonumber\\
&&-\sum_{\mathbf{k}=\mathbf{k^{'}+\mathbf{k^{''}}}}e^{-i\lambda_{Z}}\Lambda_{k^{''}}^{k^{'}}J_{k^{'}}\delta L_{k^{'}}\delta H_{k^{''}}.
\label{nonliear grokinitic eq. of ZFZS}
\end{eqnarray}
Noting the $v_{ts}/\left(qR\right)\gg\omega_{Z}$ ordering for both electrons and ions, and taking the dominant flux surface averaged quantities
$\overline{\left(\cdot\cdot\cdot\right)}\equiv\intop_{0}^{2\pi}\left(\cdot\cdot\cdot\right)d\theta/\left(2\pi\right)$,
the linear particle response to $\Omega_{Z}$ can be straightforwardly derived from the linear term of equation (\ref{nonliear grokinitic eq. of ZFZS}) as
\begin{eqnarray}
\overline{\delta H_{Z,e}^{L}} & \simeq & -\frac{e}{T_{e}}F_{Me}\left(\delta\phi-\frac{v_{\parallel}}{c}\delta A\right)_{Z},\label{electron linear response of Z}\\
\overline{\delta H_{Z,i}^{L}} & = & \frac{e}{T_{i}}F_{Mi}J_{Z}\left|\theta_{Z}\right|^{2}\left(\delta\phi-\frac{v_{\parallel}}{c}\delta A\right)_{Z},\label{ion linear response of Z}
\end{eqnarray}
where $\theta_{Z}\equiv\overline{e^{-i\lambda_{Z}}}$. Meanwhile, noting the linear particle responses to TAE given in equations (\ref{electron linear response of 0}) and (\ref{ion linear response of 0}), the nonlinear particle response to $\Omega_{Z}$ can be derived as
\begin{eqnarray}
\overline{\delta H_{Z,e}^{NL}} & = & -\frac{c}{B_{0}}k_{\theta 0}\frac{e}{T_{e}}F_{Me}\left(\frac{\omega_{*e,0}^{t}}{\omega_{0}^{2}}-\frac{k_{\parallel 0}v_{\parallel}}{\omega_{0}^{2}}\right)\partial_{r}\left|\delta\phi_{0}\right|^{2},\label{nonlinear electron response to ZF}\\
\overline{\delta H_{Z,i}^{NL}} & = & \frac{c}{B_{0}}\left|\theta_{Z}\right|^{2}k_{\theta 0}\frac{e}{T_{i}}J_{0}^{2}F_{Mi}\frac{\omega_{*i,0}^{t}}{\omega_{0}^{2}}\partial_{r}\left|\delta\phi_{0}\right|^{2},\label{nonlinear ion response to ZF}
\end{eqnarray}
where $\partial_{r}=ik_{r Z}$ is the operator for radial derivative. It is reasonable to conclude from equations (\ref{nonlinear electron response to ZF}) and (\ref{nonlinear ion response to ZF}) that, $\partial_{r}\left|\delta\phi_{0}\right|^{2}\sim ik_{r Z}\left|\delta\phi_{0}\right|^{2}$, corresponding to the fine scale ZS generation by weakly/moderately ballooning TAE \cite{ZQiuNF2017}. 
Substituting the particle responses to $\Omega_{Z}$ into the quasi-neutrality condition, nonlinear equation of zonal scalar potential generation by TAE in nonuniform plasmas can be derived as  
\begin{equation}
\begin{aligned}\delta\phi_{Z}\ =\ \frac{c}{B_{0}} & k_{\theta 0}\frac{\omega_{*i,0}}{\omega_{0}^{2}}\left[\frac{\eta_{i}}{\chi_{iZ}}\left\langle \frac{F_{Mi}}{n_{0}}J_{Z}\left|\theta_{Z}\right|^{2}J_{0}^{2}\right.\right.\\
 & \left.\left.\times\left(\frac{v^{2}}{v_{ti}^{2}}-\frac{3}{2}\right)\right\rangle-1\right]\partial_{r}\left|\delta\phi_{0}\right|^{2}.
\end{aligned}
\label{ZFZS generated by TAE 1}
\end{equation}
Here, $\chi_{iZ}\equiv1-\left\langle F_{Mi}/n_{0}J_{Z}^{2}\left|\theta_{Z}\right|^{2}\right\rangle\approx1.6k_{r Z}^{2}\rho_{ti}^{2}q^{2}/(2\sqrt{\epsilon})$
is the well known neoclassical inertia enhancement dominated by trapped
particle contribution \cite{MRosenbluthPRL1998}, and $\rho_{ti}\equiv v_{ti}/\omega_{ci}$ is the ion Larmor radius defined by ion thermal velocity. If we neglect the FLR effects by taking $J_{Z}\left(k_{r Z}\rho_{i}\right)\simeq1$, noting that, $\left\langle F_{Mi}/n_{0}\left|\theta_{Z}\right|^{2}\left(v^{2}/v_{ti}^{2}-3/2\right)\right\rangle =-\chi_{iZ}$,  equation (\ref{ZFZS generated by TAE 1}) can be explicitly written as
\begin{eqnarray}
\delta\phi_{Z} & = & -\frac{c}{B_{0}}k_{\theta 0}\frac{\omega_{*pi,0}}{\omega_{0}^{2}}\partial_{r}\left|\delta\phi_{0}\right|^{2}.\label{Eq. ZFZS generated by TAE}
\end{eqnarray}
Here, $\omega_{*pi}\equiv\omega_{*i}+\omega_{*Ti}$ with $\omega_{*Ti}=\eta_i\omega_{*i}$. Equation (\ref{Eq. ZFZS generated by TAE}) describes ZF beat-driven by TAE, different from the spontaneous excitation process described in Ref. \citenum{LChenPRL2012}. It is noteworthy that, equation (\ref{Eq. ZFZS generated by TAE}) illustrates that the growth rate of the beat-driven zonal scalar potential is twice that of the TAE instantaneous growth rate, and its magnitude being proportional to the intensity of the TAE. This is a typical feature of the beat-driven process \cite{HWangPOP2020, ZQiuPoP2016, YTodoNF2010}, with the crucial role played by the nonlinear ion response to ZF due to thermal ion nonuniformity. The beat-driven process is thresholdless, significantly different from the spontaneous excitation process via modulational instability of TAE, which requires a sufficiently large TAE amplitude to overcome the threshold due to frequency mismatch \cite{ZQiuNF2017}. This is the reason why the beat-driven process is universally observed in simulations \cite{YTodoNF2010, ABiancalaniJPP2020, GDongPoP2019, HWangPOP2020}.

Meanwhile, the equation for zonal current generation can be derived from the parallel component of Ampere's law \cite{LChenPRL2012}. One obtains from the zonal component of equation (\ref{parallel Ampere's law}), 
\begin{eqnarray}
\delta A_{\parallel Z}&\simeq&-\frac{c^{2}}{B_{0}}k_{\theta 0}\frac{k_{\parallel 0}}{\omega_{0}^{2}}\partial_{r}\left|\delta\phi_{0}\right|^{2},\label{zonal current}
\end{eqnarray}
with the assumption of $k_{\perp}^{2}c^{2}/\omega_{pe}^{2}\ll1$, and $\omega_{pe}^{2}=4\pi n_{0}e^{2}/m_{e}$ being the electron plasma frequency. Equations (\ref{Eq. ZFZS generated by TAE})-(\ref{zonal current}), together with nonlinear particle response to ZS given by equations (\ref{nonlinear electron response to ZF})-(\ref{nonlinear ion response to ZF}) (i.e., PSZS), will be used to study TAE effects on ITG via ZS mediation in section \ref{Effects of beat-driven ZFZS on DW linear stability}.

\section{Effects of beat-driven ZS on DW linear stability}\label{Effects of beat-driven ZFZS on DW linear stability}

Next, we consider the effects of beat-driven ZS on ITG linear stability. For simplicity, we consider the ITG to be electro-static, and the analysis follows closely that of Ref.\citenum{NChenPoP2021}. The ITG equation is derived from quasi-neutrality condition, i.e., equation (\ref{quasineutrality condition}), with the particle responses derived from nonlinear gyrokinetic equation.

For ITG with $k_{\parallel}v_{te}\gg\omega\sim\omega_{*i}\gg\omega_{Ds}$, the linear electron response to ITG is adiabatic, i.e., 
\begin{eqnarray}
\delta H_{I,e}^{L}&\approx&0;\nonumber
\end{eqnarray}
and for ion with $\omega\gg k_{\parallel}v_{ti}$, the linear ion responses to ITG can be derived as 
\begin{eqnarray}
\delta H_{I,i}^{L}\simeq \frac{e}{T_{i}}\left(1-\frac{\omega_{*i}^{t}}{\omega}\right)\left(1+\frac{v_{\parallel}k_{\parallel}}{\omega}+\frac{v_{\parallel}^{2}k_{\parallel}^{2}}{\omega^{2}}+\frac{\omega_{Di}}{\omega}\right)F_{Mi}J_{0}\delta\phi_{I}.\nonumber
\end{eqnarray}
Here, we have adopted the fluid-ion approximation in order to simplify the analysis and, thereby, illustrate more clearly the effects of ZS on ITG stabilizing. Substituting the linear particle responses to ITG into quasi-neutrality condition, the linear ITG equation can be derived as 
\begin{eqnarray}
\epsilon_{I}\delta\phi_{I} & = & 0,\label{equation of the linear ITG dispersion relation}
\end{eqnarray}
with
\begin{eqnarray}
\epsilon_{I}\ =\ 1-\frac{\omega_{*e}}{\omega}+\tau\left(1-\frac{\omega_{*pi}}{\omega}\right) \left( b_{\perp}-\frac{  k_{\parallel}^{2}v_{ti}^{2}}{2\tau\omega^{2}}-\frac{2 \omega_{di}C}{\omega}\right)  \label{linear ITG dispersion relation}
\end{eqnarray}
being the linear ITG dielectric operator, and $b_{\perp}=k_{\perp}^{2}\rho_{i}^{2}/2$. The contributions of beat-driven ZS, enter through the formally ``nonlinear'' terms, including scattering by zonal field structure (i.e., $\delta\phi_Z$ and $\delta A_{\parallel Z}$) via $J_{Z}\delta L_{Z}\delta H_{I}$ and nonlinear modification of equilibrium by PSZS via $J_{I}\delta \phi_{I}\delta H^{NL}_{Z}$ in the nonlinear gyrokinetic equation
\begin{equation}
\begin{aligned}\left(\partial_{t}+v_{\parallel}\partial_{l}+i\omega_{D}\right)\delta H_{I}^{NL}\ =\ \Lambda & \left[J_{I}\delta\phi_{I}\left(\delta H_{Z}^{L}+\delta H_{Z}^{NL}\right)\right.\\
 & \left.-J_{Z}\delta L_Z\delta H_{I}^{L}\right],
\end{aligned}
\label{nonlinear particle response to ITG}
\end{equation}
with $\Lambda=-ck_{r Z}k_{\theta I}/B_0$. Noting the $k_{\parallel}v_{te}\gg\omega\gg k_{\parallel}v_{ti}$ ordering again, we have
\begin{eqnarray}
\delta H_{I,e}^{NL} & \simeq & 0,\label{nonlinear electron response to ITG}
\end{eqnarray}
and 
\begin{equation}
\begin{aligned}\delta H_{I,i}^{NL}\ =\ \frac{e}{T_{i}}F_{Mi} & \left[\left(\left|\theta_{Z}\right|^{2}-1+\frac{\omega_{*i}^{t}}{\omega}\right)\frac{\omega_{*pi}}{\omega}\right.\\
 & \left.-\frac{\omega_{*i}^{t}}{\omega}\left|\theta_{Z}\right|^{2}\right]\delta\hat{\phi}_{0}^{2}\delta\phi_{I}.
\end{aligned}
\label{nonlinear ion response to ITG}
\end{equation}
Here, $\delta\hat{\phi}^2_{0}=\left| ck_{\theta 0}/\left(B_{0}\omega_{0}\right)\right|^2\partial^2_{r}\left|\delta\phi_{0}\right|^{2}$. Note that $\delta\hat{\phi}_0$ is related to the normalized Doppler shift induced by radial electric field of TAE,  and is also related to the beat-driven ZF induced shearing rate. In deriving equations (\ref{nonlinear electron response to ITG}) and (\ref{nonlinear ion response to ITG}), we neglected the FLR effects by taking $J_{Z}\left(k_{r Z}\rho_{i}\right)\simeq1$, and the expressions of the   beat-driven ZS given by equations (\ref{nonlinear electron response to ZF}), (\ref{nonlinear ion response to ZF}), (\ref{Eq. ZFZS generated by TAE}) and (\ref{zonal current}) are used. Substituting
equations (\ref{nonlinear electron response to ITG}) and (\ref{nonlinear ion response to ITG}) into the quasi-neutrality condition, the equation describing nonlinear modulation of ITG by the beat-driven ZS can be written as
\begin{equation}
\frac{n_{0}e^{2}}{T_{e}}\left[ \epsilon_{I}+\tau\frac{\omega_{*i}}{\omega}\left(1-\frac{\omega_{*pi}}{\omega}\right)\delta\hat{\phi}_{0}^{2}\right] \delta\phi_{I}\ =\ 0.\label{beat-driven ZFZS on ITG}
\end{equation}
Combining equations (\ref{linear ITG dispersion relation}) and (\ref{beat-driven ZFZS on ITG}), one then has the ITG dispersion relation in the WKB limit
\begin{equation}
\begin{aligned} & \left[\frac{\omega}{\tau\left(\omega-\omega_{*pi}\right)}+\frac{\omega_{*i}}{\omega-\omega_{*pi}}+b_{\perp}-\frac{k_{\parallel}^{2}v_{ti}^{2}}{2\omega^{2}}\right.\\
 & \left.-\frac{2\omega_{di}C}{\omega}+\frac{\omega_{*i}}{\omega}\delta\hat{\phi}_{0}^{2}\right]\delta\phi_{I}\ =\ 0.
\end{aligned}
\label{ITG WKB D.R.}
\end{equation}
Here, operators $\omega$, $b_{\perp}$ and $k_{\parallel}$ without subscripts $k=0, Z$ represent ITG.
The first five terms of equation (\ref{ITG WKB D.R.}) constitute the linear ITG dispersion relation, which deviates slightly from that in Ref. \citenum{NChenPoP2021}, primarily due to the inclusion of both density and temperature nonuniformities in the present analysis, while Ref. \citenum{NChenPoP2021} took the flat density gradient limit to focus on the effects of ion temperature gradient, following the early analysis of Ref. \citenum{CChengPoF1980}. Moreover, the first four terms correspond to the adiabatic electron response, ion $\mathbf{E}\times\mathbf{B}$ drift, the FLR effect (polarization), and the the parallel compressibility, respectively. The fifth term accounts for the magnetic drift, which is unique to toroidal configurations and leads to the coupling of adjacent poloidal harmonics. The last term introduces a nonlinear modification attributed to the beat-driven ZS, and is the main novel contribution of the present work. 
Additionally, in this context, the final term represents a nonlinear correction that takes into account the combined contributions of ZF, ZC, and PSZS. The contribution of each of the three components to the nonlinear ITG dispersion relation will be presented in the Appendix \ref{Appendix}.

Noting $k_{\perp}^{2}=k_{\theta}^{2}-\partial^{2}/\partial r^{2}$, and introducing the Fourier transform $\Phi\left(\eta\right)=\intop \Phi (nq-m)\exp\left(-i\eta(nq-m)\right)d(nq-m)$ with $nq-m$ being the normalized distance to the mode rational surface and $\Phi(nq-m)$ being the poloidal harmonic of ITG, the ITG eigenmode equation in ballooning space  can be derived as
\begin{equation}
\begin{aligned} & \frac{d^{2}\Phi\left(\eta\right)}{d\eta^{2}}+q^{2}\Omega^{2}b\left[-\frac{\tau}{\left(1+\tau\Omega\epsilon_{pi}^{1/2}\right)\left(1+\eta_{i}\right)\epsilon_{pi}^{1/2}}\right.\\
 & +\frac{\tau\Omega}{1+\tau\Omega\epsilon_{pi}^{1/2}}+b\left(1+\hat{s}^{2}\eta^{2}\right)+\frac{2}{\Omega}\left(\cos\eta+\hat{s}\eta\sin\eta\right)\\
 & \left.-\frac{1}{\Omega\epsilon_{pi}\left(1+\eta_{i}\right)}\delta\hat{\phi}_{0}^{2}\right]\Phi\left(\eta\right)\ =\ 0,
\end{aligned}
\label{ITG eigenmode equation in ballooning space}
\end{equation}
where $\hat{s}\equiv r\left(\partial q/\partial r\right)/q$ denotes the magnetic shear,  $\Omega\equiv\omega/\left(\tau\sqrt{\omega_{*pi}\omega_{di}}\right)$,
$b\equiv\tau b_{\theta}/\sqrt{\epsilon_{pi}}$, and $\epsilon_{pi}\equiv L_{pi}/R$.
Equation (\ref{ITG eigenmode equation in ballooning space}) is the paradigmatic one-dimensional eigenvalue equation with the potential well contributed by two components: a slowly varying parabolic well stemming from the FLR effect $(1+\hat{s}^{2}\eta^{2})$, and a rapidly oscillating periodic well due to magnetic curvature $(\cos\eta+\hat{s}\eta\sin\eta)$, while the beat-driven ZS by TAE enter the ITG eigenmode equation by modifying the potential well depth while not affecting the parity or periodicity. These distinct potential wells result in two principal ITG eigenmode branches: the slab branch sensitive to the slow parabolic variation $(1+\hat{s}^{2}\eta^{2})$, and the toroidal branch due to the rapid periodic fluctuations $(\cos\eta+\hat{s}\eta\sin\eta)$. In section \ref{Analytical and numerical results}, following the analysis of Refs. \citenum{PGuzdarPoF1983} and \citenum{LChenPoFB1991}, we assess the impact of ZS on ITG characteristics, including frequency, growth rate, and mode structure, in the short- and long-wavelength regimes, respectively. The analytical results are then compared with numerical solution of equation (\ref{ITG eigenmode equation in ballooning space}) using the eigenmatrix method. In equation (\ref{ITG eigenmode equation in ballooning space}), we also note that $\delta\hat{\phi}^2_{0}=\left|ck_{\theta 0}/(B_{0}\omega_{0})\right|^2\partial^2_{r}\left|\delta\phi_{0}\right|^{2}\sim-\left|k_{r Z}/k_{\parallel 0}\right|^{2}\left|\delta B_{r}/B_0\right|^{2}$, assuming the TAE is close to ideal MHD condition. Taking typical tokamak parameters, i.e.,  $k_{r Z}\sim1/\rho_E$, $k_{\parallel 0}\sim1/(2qR)$, and $\left|\delta B_{r}/B_0\right|^2\leq 2.5\times 10^{-7}$ for TAE fluctuations level expected in reactors \cite{WHeidbrinkPRL2007}, we then find $\delta\hat{\phi}^2_{0}\sim \mathcal{O}(10^{-1})$, and $e\delta\phi_Z/T_i\leq 5\times 10^{-4}$.

\section{Analytical and numerical results}\label{Analytical and numerical results}

In this section, we follow the theoretical approach of Refs.\citenum{PGuzdarPoF1983} and \citenum{LChenPoFB1991}, and investigate the impact of TAE-driven-ZS on ITG linear stability in both short- and long-wavelength limits, corresponding to strong and moderate ballooning cases, respectively. The two limiting parameter regimes, can be studied by taking $b\sim\mathcal{O}\left(1\right)$ and $b\ll1$, respectively, which is given \textit{a prior}, and will be shown below.

\subsection{Short-wavelength limit}

\begin{figure}[]
    \centering
    \includegraphics[scale=0.085]{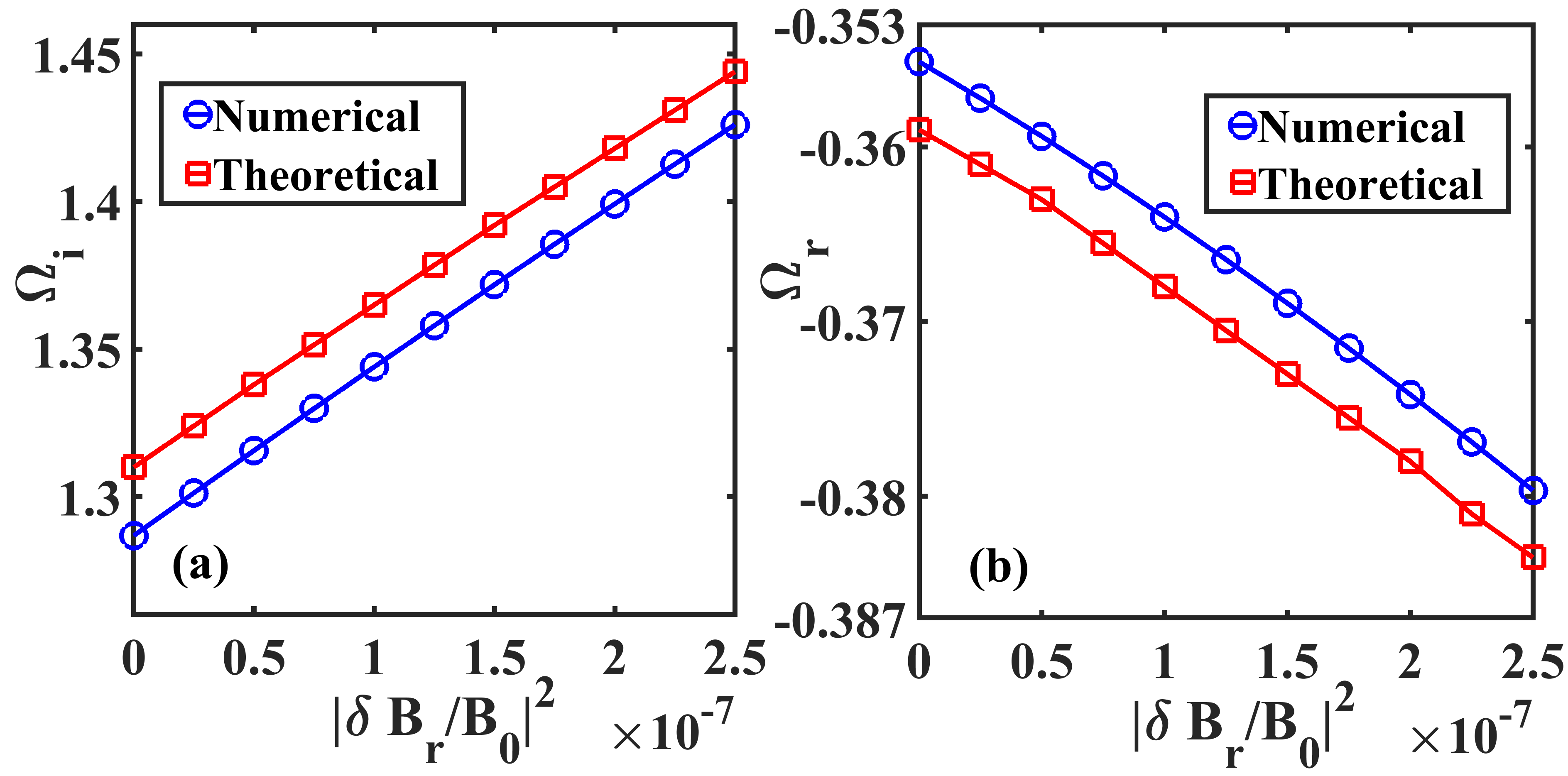}
    \caption{The dependence of normalized ITG growth rate $(a)$ and real frequency $(b)$ vs TAE amplitude in short-wavelength limit. The squares represent the theoretical result given by equation (\ref{dispersion relation in short-wavelength limit}) while circles are numerical results of equation (\ref{ITG eigenmode equation in ballooning space}).
    Here, $\eta_{i}=2$, $\epsilon_{pi}=0.06$, $q=1$ and $b=1$.}\label{short} 
\end{figure}

In the short-wavelength limit, i.e., $b\sim\mathcal{O}\left(1\right)$, the parallel mode structure of ITG is strongly localized in ballooning space\cite{PGuzdarPoF1983}. Assuming that the ITG eigenmode is localized around $\eta=0$ in the ballooning space, in which strong coupling approximation can be adopted by taking $\cos\eta\approx1-\eta^{2}/2$ and $\sin\eta\approx\eta$, and the ITG eigenmode equation becomes
\begin{equation}
\begin{aligned} & \frac{d^{2}\Phi\left(\eta\right)}{d\eta^{2}}+q^{2}\Omega^{2}b\left[-\frac{\tau}{\left(1+\tau\Omega\epsilon_{pi}^{1/2}\right)\left(1+\eta_{i}\right)\epsilon_{pi}^{1/2}}\right.\\
 & +\left.\frac{\tau\Omega}{1+\tau\Omega\epsilon_{pi}^{1/2}}+b+\frac{2}{\Omega}-\frac{1}{\Omega\epsilon_{pi}\left(1+\eta_{i}\right)}\delta\hat{\phi}_{0}^{2}\right.\\
 & \left.+\left(b\hat{s}^{2}+\frac{2\hat{s}-1}{\Omega}\right)\eta^{2}\right]\Phi\left(\eta\right)\ =\ 0,
\end{aligned}
\label{the eigenmode equation}
\end{equation}
which can be rewritten as a standard Weber equation with the most unstable ground eigenmode given by $\delta\phi=\exp\left(-\sigma\eta^{2}\right)$, with
\begin{equation}
\begin{aligned}\sigma= & \frac{q^{2}\Omega^{2}b}{2}\left[-\frac{\tau}{\left(1+\tau\Omega\epsilon_{pi}^{1/2}\right)\left(1+\eta_{i}\right)\epsilon_{pi}^{1/2}}\right.\\
 & \left.+\frac{\tau\Omega}{1+\tau\Omega\epsilon_{pi}^{1/2}}+b+\frac{2}{\Omega}-\frac{1}{\Omega\epsilon_{pi}\left(1+\eta_{i}\right)}\delta\hat{\phi}_{0}^{2}\right].
\end{aligned}
\end{equation}
The half width of the ground-state eigenmode in ballooning space is proportional to $1/\sqrt{b}$, and consequently, it is proportional to $\sqrt{b}$ in real space, consistent with the strong coupling approximation. 
The corresponding dispersion relation is
\begin{equation}
\begin{aligned} & q^{2}\Omega^{2}b\left[-\frac{\tau}{\left(1+\tau\Omega\epsilon_{pi}^{1/2}\right)\left(1+\eta_{i}\right)\epsilon_{pi}^{1/2}}\right.\\
 & \left.+\frac{\tau\Omega}{1+\tau\Omega\epsilon_{pi}^{1/2}}+b+\frac{2}{\Omega}-\frac{1}{\Omega\epsilon_{pi}\left(1+\eta_{i}\right)}\delta\hat{\phi}_{0}^{2}\right]^{2}\\
 & +\left(b\hat{s}^{2}+\frac{2\hat{s}-1}{\Omega}\right)\ =\ 0.
\end{aligned}
\label{dispersion relation in short-wavelength limit}
\end{equation}
The dependence of ITG growth rate and real frequency on the amplitude of ZS is solved from the theoretical dispersion relation equation (\ref{dispersion relation in short-wavelength limit}), which are then compared with the numerical solution of equation (\ref{ITG eigenmode equation in ballooning space}), as shown in FIG. \ref{short}. 

\begin{figure}[]
    \centering
    \includegraphics[scale=0.088]{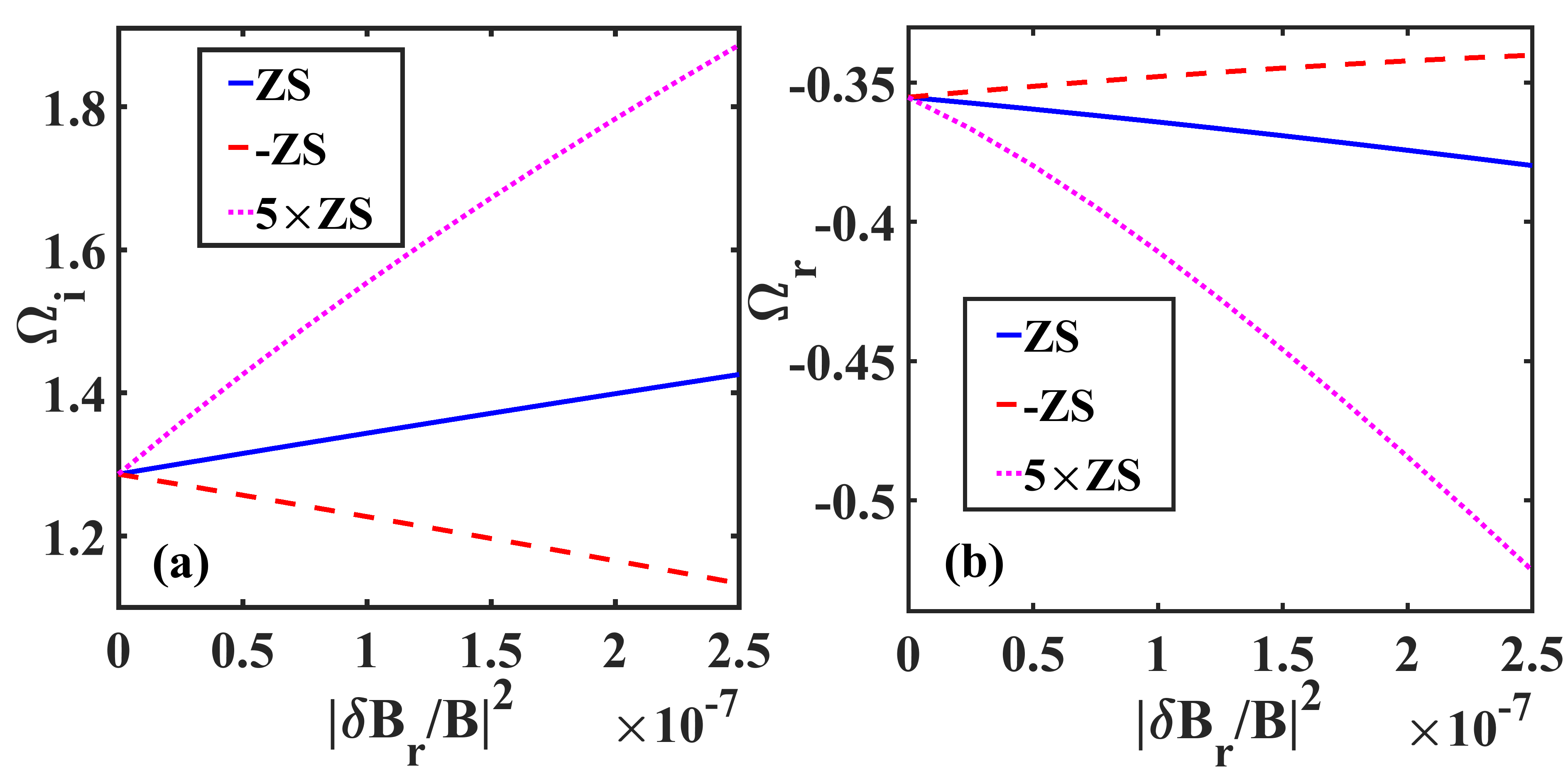}
    \caption{The dependence of normalized ITG growth rate $(a)$ and real frequency $(b)$ vs TAE amplitude in short-wavelength
    limit. 
    Here, $\eta_{i}=2$, $\epsilon_{pi}=0.06$, $q=1$ and $b=1$.}\label{short_wavelength_limit_with ZS} 
\end{figure}

Good agreement between analytical and numerical results are obtained. It is found that, the ITG real frequency as well as growth rate change only slightly with TAE amplitude with the typical tokamak parameters regime, i.e., $\left|\delta B_{r}/B_0\right|^2\leq 2.5\times10^{-7}$. More importantly, the ITG growth rate increases with TAE amplitude, suggesting the ZS beat driven by TAE, has only weakly destabilizing effect on ITG stability, contrary to the general speculation. 
In particular, the influence of ZS on the ITG in our work is significantly less pronounced than the results reported in Ref. \citenum{NChenPoP2021}, where effects of radial electric field from electro-static ZFZF on ITG stability are investigated. This discrepancy primarily stems from the coefficient of the nonlinear term  in our equation (\ref{ITG eigenmode equation in ballooning space}) being substantially smaller than the corresponding coefficient in equation (7) of Ref. \citenum{NChenPoP2021}. 
In fact, even at the maximum TAE amplitude adopted, the amplitude of the ZS beat-driven by TAE is also quite small, i.e., $e\delta\phi_Z/T_i\leq5\times10^{-4}$, so it is reasonable that the impact of the TAE beat-driven ZS on the ITG linear stability is correspondingly weak.
If the coefficient of the nonlinear term is artificially enlarged by a factor of five, as depicted by the magenta dotted line in FIG. \ref{short_wavelength_limit_with ZS}, the destabilizing effect of TAE beat-driven ZS on ITG becomes markedly evident.  
Besides, in present work, linear growth rate of ITG increase with TAE amplitude, which is in contrast to the numerical results in Ref. \citenum{NChenPoP2021}. The primary reason is attributed to the difference in the sign of the nonlinear terms, which is opposite to that  in Ref. \citenum{NChenPoP2021}. As the sign of the nonlinear term in Equation (\ref{ITG eigenmode equation in ballooning space}) is artificially reversed, as illustrated by the red long dashed line in FIG. \ref{short_wavelength_limit_with ZS}, the ITG growth rate decreases with TAE amplitude,  similar to the results of Ref. \citenum{NChenPoP2021}. However, the underlying physical mechanisms remain elusive and is worthy of further investigation.

\begin{figure}[]
    \centering
    \includegraphics[scale=0.085]{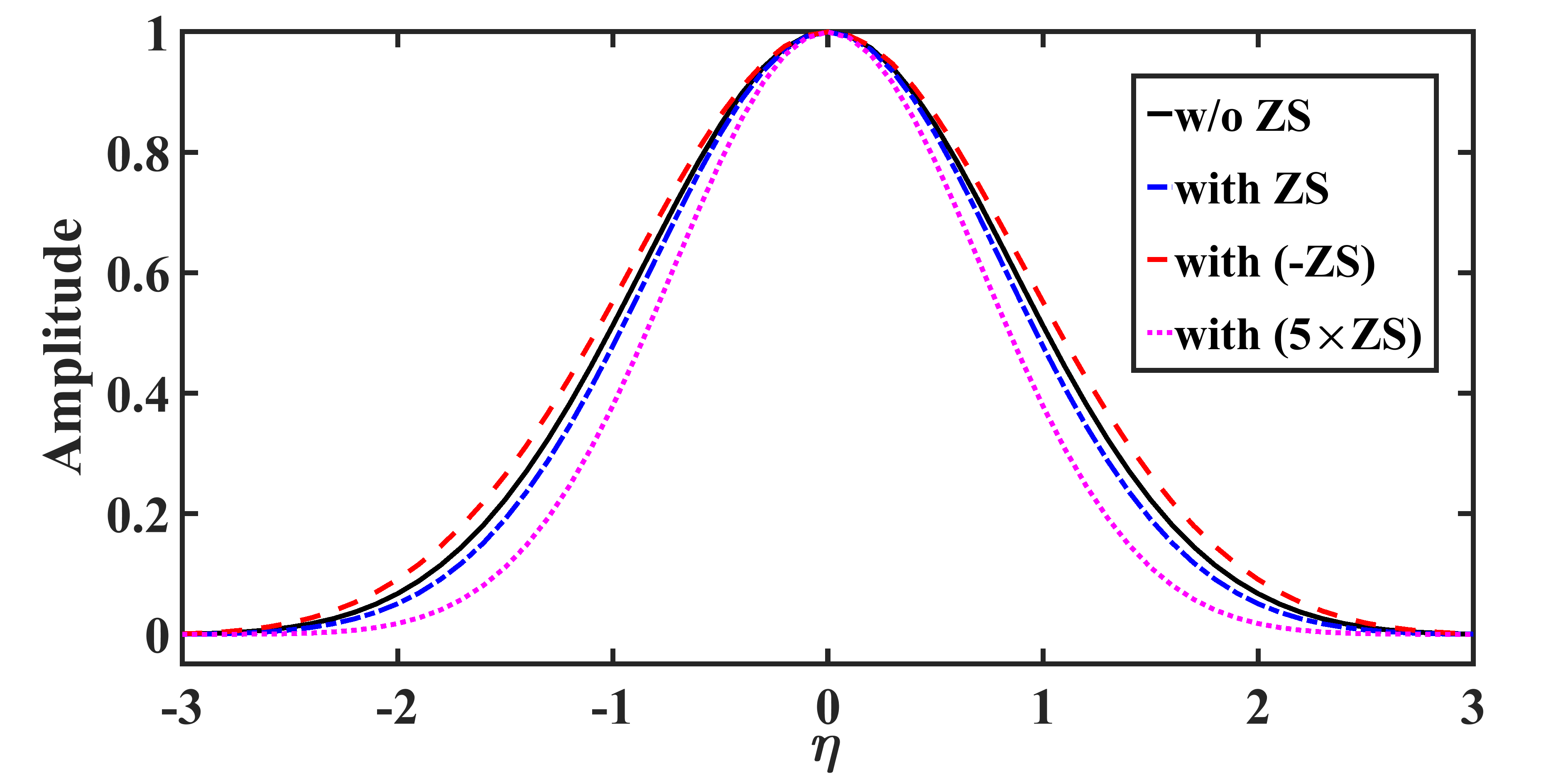}
    \caption{The mode structures of the most unstable mode with different ZS values. Here, $\epsilon_{pi}=0.06$, $q=1$ and $b=1$.}\label{Mode structure with ZS} 
\end{figure}

\begin{figure}[]
    \centering
    \includegraphics[scale=0.085]{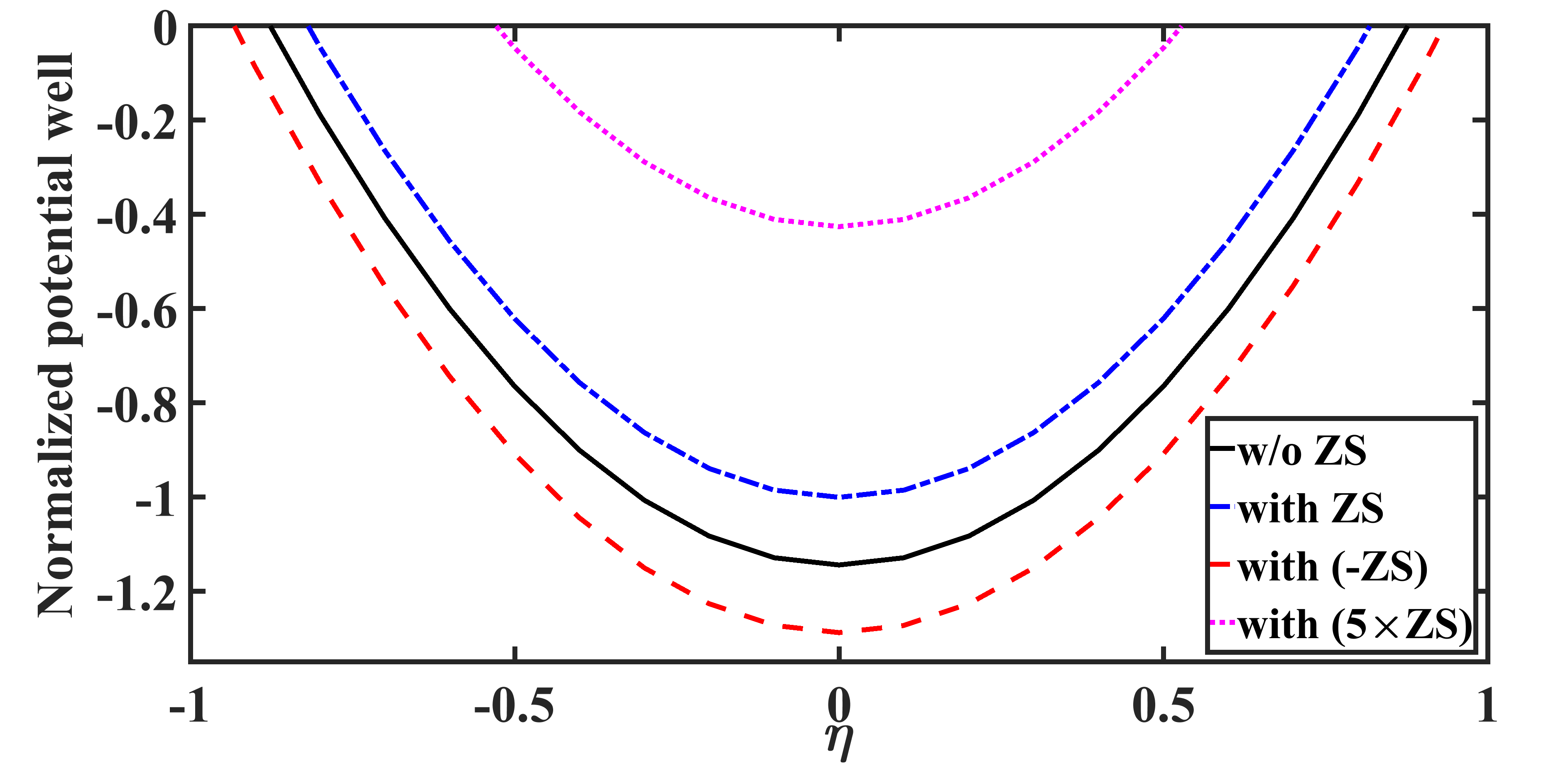}
    \caption{Potential wells with different values of ZS. Here, $\epsilon_{pi}=0.06$, $q=1$ and $b=1$.}\label{potential well with ZS} 
\end{figure}

The corresponding mode structures of the most unstable mode are shown in FIG. \ref{Mode structure with ZS}, where the    blue dotdash curve represents the case with  $\left|\delta B_{r}/B_0\right|^2=2.5\times10^{-7}$, red-dashed  and  magenta dotted curves represent  the cases with the nonlinear term artificially changed sign and increased by 5 times, respectively; and the black curve represents  the linear mode structure.
Clearly, the TAE beat-driven ZS  reduces the half-width of the most unstable ITG mode structure. However, even at the maximum amplitude of TAE, the impact of the TAE thermally driven ZS on the ITG mode structure remains minimal. 
The primary reason is that the TAE beat-driven ZS slightly reduces the depth of the potential well, thereby decreasing the distance between the turning points of the potential well and localizing the ITG mode within a narrower range of stronger drive at the bad curvature region, as shown in FIG. \ref{potential well with ZS}, and this is confirmed by the magenta dotted curve and red dashed curve, where the nonlinear term is artificially changed sign or enlarged by a factor.
Additionally, the mode structure peaks at $\eta=0$ and the even symmetry remains unbroken, a result of the even-symmetric modulation introduced by the beat-driven ZS.

\subsection{Long-wavelength limit}

For typical tokamak plasmas, strong coupling approximation is usually a rough approximation. In more general cases, long wavelength limit with $b\ll1$  is satisfied \cite{LChenPoFB1991}. In the long-wavelength limit, there are two branches, i.e., the toroidal
branch and the slab branch. We are more concerned about the toroidal branch \cite{LChenPoFB1991}, 
which is characterized by fast variation over connection length scale ($\eta\sim\mathcal{O}\left(1\right)$) and a superimposed slowly varying
envelope over secular scale. Following the analysis of Ref. \citenum{LChenPoFB1991}, taking
$\Phi\left(\eta\right)=C_{0}\left(\eta_{1}\right)\cos\eta/2+S_{0}\left(\eta_{1}\right)\sin\eta/2$
with $\eta_{1}\equiv\hat{\epsilon}\eta$  and $\hat{\epsilon}=b^{1/3}$
denoting slow variation in $\eta$, the eigenmode equations can be
derived from vanishing coefficients of the linearly independent bases $\sin\eta/2$ and $\cos\eta/2$,
\begin{eqnarray}
\frac{dS_{0}}{d\eta_{1}} &+ & \left[\frac{q^{2}b^{2/3}\Omega^{3}\tau}{1+\tau\Omega\epsilon_{pi}^{1/2}}-\frac{q^{2}b^{2/3}\Omega^{2}\tau}{\left(1+\tau\Omega\epsilon_{pi}^{1/2}\right)\left(1+\eta_{i}\right)\epsilon_{pi}^{1/2}}-\frac{1}{4b^{1/3}}\right.\nonumber\\
 && \left.-\frac{q^{2}b^{2/3}\Omega}{\epsilon_{pi}\left(1+\eta_{i}\right)}\delta\hat{\phi}_{0}^{2}\right]C_{0}+q^{2}\Omega b^{1/3}\hat{s}\eta_{1}S_{0}=0,
\label{equation of S0}\\
 \frac{dC_{0}}{d\eta_{1}} &- & \left[\frac{q^{2}b^{2/3}\Omega^{3}\tau}{1+\tau\Omega\epsilon_{pi}^{1/2}}-\frac{q^{2}b^{2/3}\Omega^{2}\tau}{\left(1+\tau\Omega\epsilon_{pi}^{1/2}\right)\left(1+\eta_{i}\right)\epsilon_{pi}^{1/2}}-\frac{1}{4b^{1/3}}\right.\nonumber\\
 && \left.-\frac{q^{2}b^{2/3}\Omega}{\epsilon_{pi}\left(1+\eta_{i}\right)}\delta\hat{\phi}_{0}^{2}\right]S_{0}-q^{2}\Omega b^{1/3}\hat{s}\eta_{1}C_{0}=0.
\label{equation of C0}
\end{eqnarray}
Equation \eqref{equation of S0} and \eqref{equation of C0} can be cast into a Weber equation for $C_{0}$ and $S_{0}$, which then yields the dispersion relation for the most unstable ground eigenstate  
\begin{equation}
\begin{aligned}\Omega^{3}\ -\ & \Omega^{2}\left[\frac{1}{\left(1+\eta_{i}\right)\epsilon_{pi}^{1/2}}-\frac{1}{\left(1+\eta_{i}\right)\epsilon_{pi}^{1/2}}\delta\hat{\phi}_{0}^{2}\right.\\
 & \left.\times\left(1+\frac{1}{\tau\Omega\epsilon_{pi}^{1/2}}\right)\right]= \frac{1}{4bq^{2}\tau}\left(1+\tau\Omega\epsilon_{pi}^{1/2}\right).
\end{aligned}
\label{dispersion relation in long-wavelength limit}
\end{equation}

The dispersion relation is similar to corresponding linear result \cite{LChenPoFB1991}, with the term proportional to $\delta\hat{\phi}_{0}^{2}$ originating
from the contribution of beat-driven ZS. 

\begin{figure}[]
    \centering
    \includegraphics[scale=0.085]{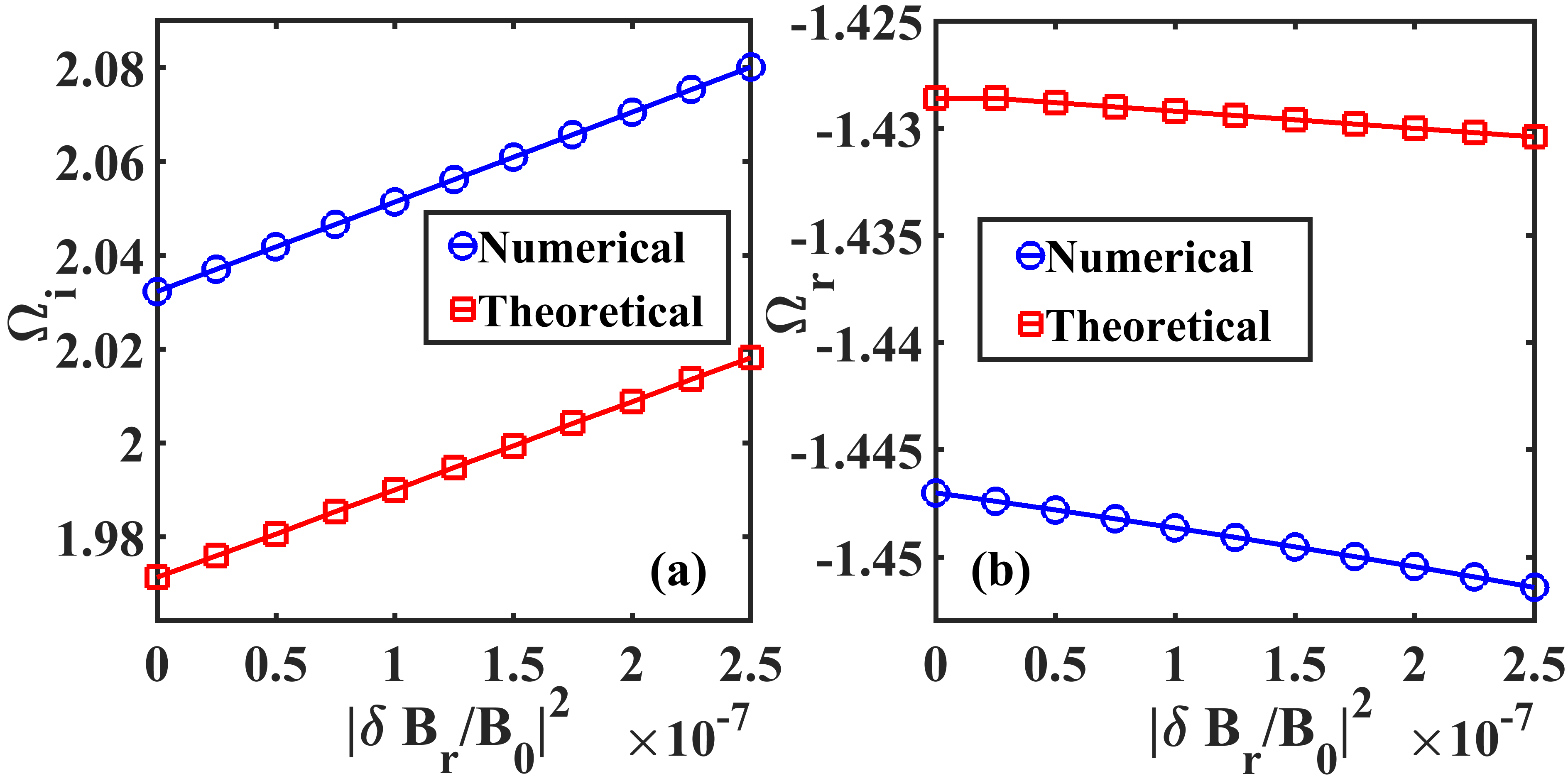}
    \caption{The dependence of normalized ITG growth rate $(a)$ and real frequency $(b)$ vs the TAE amplitude in long-wavelength limit. The squares represent the theoretical result given by equation (\ref{dispersion relation in long-wavelength limit}) while circles are numerical results of equation (\ref{ITG eigenmode equation in ballooning space}).
    Here, $\eta_{i}=2$, $\epsilon_{pi}=0.06$, $q=1$ and $b=0.01$.}\label{long} 
\end{figure}

\begin{figure}[]
    \centering
    \includegraphics[scale=0.085]{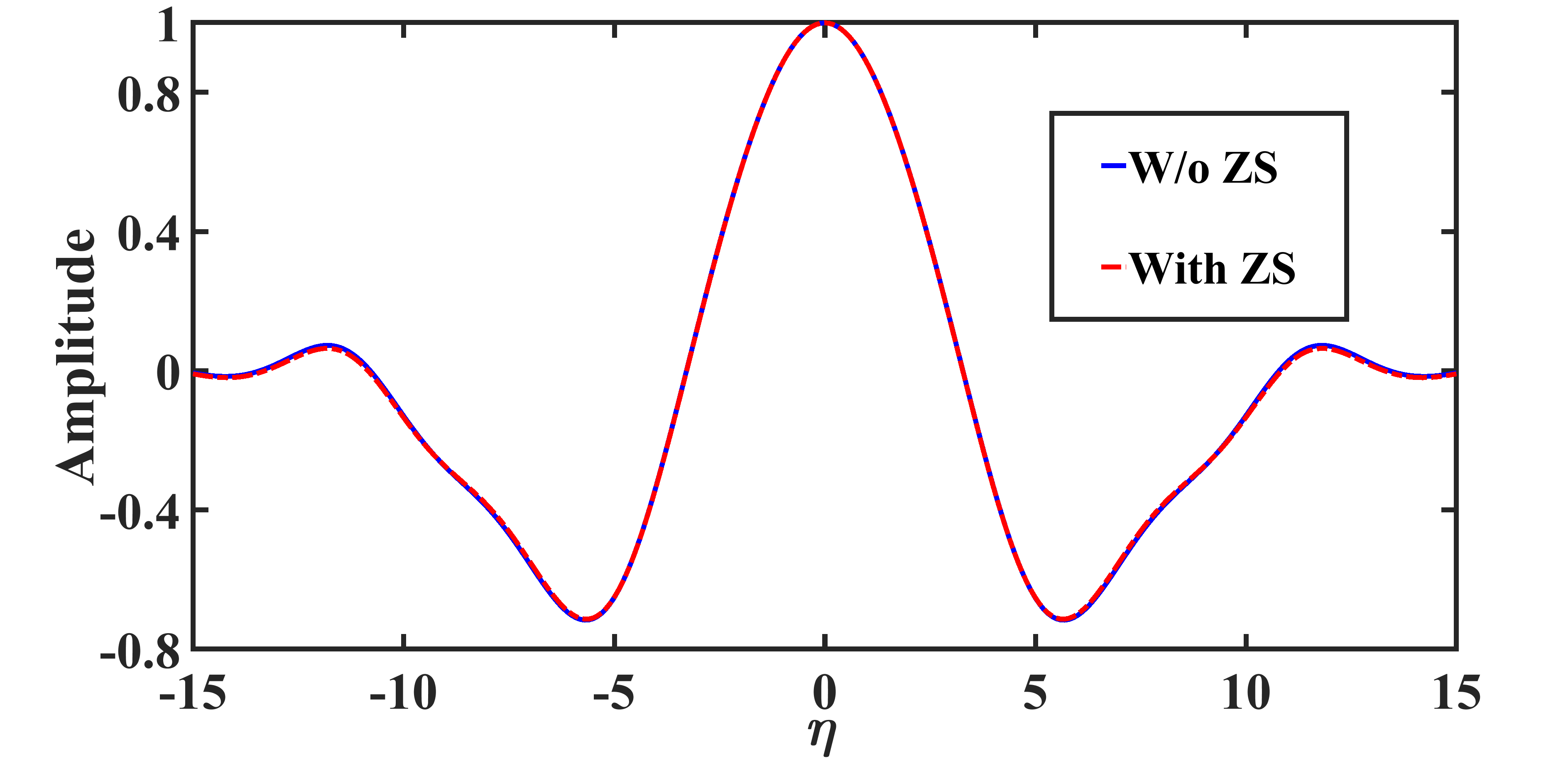}
    \caption{The mode structure of the most unstable mode. Here, $\epsilon_{pi}=0.06$, $q=1$ and $b=0.01$.}\label{mode structure in long} 
\end{figure}

Following the same procedures, the dependence of the ITG growth rate and real frequency on beat-driven ZS in the long-wavelength limit is illustrated in FIG. \ref{long}.
Analytical and numerical results indicate that in the long-wavelength limit, the growth rate as well as real frequency of the toroidal branch vary  in a similar manner to the  results of  the short-wavelength limit. 
That is, the ITG growth rate slightly increases with TAE amplitude, indicating  that ZS beat-driven by TAE has only  a weakly destabilizing effect on ITG stability due to the relatively small amplitude of the ZS, as well as its phase. The analysis methods are similar to  those used in the short-wavelength limit, and thus, will not be repeated here.
It is noteworthy that, an artificially small $b=0.01$ is adopted here to separate different scales for analytical progress, due to the weak dependence of $\hat{\epsilon}\propto b^{1/3}$, although this is not the most relevant parameter regime for ITG stability.
The mode structure in the long-wavelength limit is presented in FIG. \ref{mode structure in long}, which indicates clearly that the impact of the TAE beat-driven ZS on the most unstable mode is negligible.

\section{Summary and discussion}\label{Summary and discussion}

In this work, the indirect nonlinear interaction between toroidal Alfv\'en eigenmode (TAE) and ion temperature gradient mode (ITG) is investigated using nonlinear gyrokinetic theory and ballooning formalism, to understand the thermal plasma confinement in the presence of energetic particles (EPs) and the associated electromagnetic oscillations. More specifically, the local  stability of ITG in the presence of zonal structures (ZS) beat-driven by finite amplitude TAE is analyzed. The governing ITG eigenmode equation in the ballooning space is derived and solved both analytically and numerically, and it is found that, for typical reactor parameters and TAE fluctuation level, the ZS beat-driven by TAE has only weakly destabilizing effects on ITG stability in both short- and long-wavelength limits, contrary to the previous speculations. This is due to the relatively weak ZS generation in the parameter region, i.e., $|e\delta\phi_Z/T_i|\sim O(10^{-4})$, partly due to the scale separation between TAE and ITG. 

In the present analysis, the contributions  of zonal field structure, including zonal flow (ZF), zonal current (ZC), and phase space zonal structure (PSZS) are accounted for on the same footing, and only the overall results are analysed. The respective contributions of zonal field structure and PSZS to the ITG dispersion relation in the WKB limit are provided in the Appendix. However, their respective  contribution to the ITG stability are not analysed, as the structure of the corresponding eigenmode equation in the ballooning space is significantly modified, due to the non-vanishing nonlinear electron response to ITG, which makes the comparison to the present results not relevant. 

The major assumptions made in the present analysis, include 1. the ZS are beat-driven by TAE, and 2. only the local stability of ITG is investigated. It is noteworthy that,   in a previous study, it is found that, the direct scattering of electron drift wave (eDW) by finite amplitude ambient TAE has negligible effects on eDW stability \cite{LChenNF2023}. The present work, together with Ref. \citenum{LChenNF2023}, thus, ruled out most channels for the regulation of ITG by ambient electromagnetic oscillations driven by EPs, such as TAE. The potentially effective channels to stabilize DW turbulence, including the spontaneously excited ZS by TAE via modulational instability, as well as DW radial envelope modulation by ZS, are ongoing work of the team, and will be reported in future publications.

\section*{Acknowledgement}
This work was supported by the Strategic Priority Research Program of Chinese Academy of Sciences under Grant No. XDB0790000, the National Science Foundation of China under Grant Nos. 12275236 and 12261131622, and  Italian Ministry for Foreign Affairs and International Cooperation Project under Grant  No. CN23GR02. This work was also supported by the EUROfusion Consortium, funded by the European Union via the Euratom Research and Training Programme (Grant Agreement No. 101052200 EUROfusion). The views and opinions expressed are, however, those of the author(s) only and do not necessarily reflect those of the European Union or the European Commission. Neither the European Union nor the European Commission can be held responsible for them.

\appendix
\section{Nonlinear contributions of ZFS and PSZS to ITG}\label{Appendix}

Here, we will give the respective contributions of zonal field structure (ZFS, i.e., $\delta\phi_Z$ and $\delta A_{\parallel Z}$) and PSZS (i.e., $\delta H^{NL}_{Z}$) on the ITG linear stability. In the nonlinear gyrokinetic equation for nonlinear particle response to ITG
\begin{equation}
\begin{aligned}
\left(\partial_{t}+v_{\parallel}\partial_{l}+i\omega_{D}\right)\delta H_{I}^{NL}&\ =\ \Lambda\left[J_{I}\delta\phi_{I}\left(\delta H_{Z}^{L}+\delta H_{Z}^{NL}\right)\right.\\
& \left. -J_{Z}\left(\delta\phi_{Z}-\frac{v_{\parallel}}{c}\delta A_{\parallel Z}\right)\delta H_{I}^{L}\right],
\end{aligned}
\label{nonlinear particle response to ITG in A}
\end{equation}
the first and third term in the square bracket  on the right hand side are contributions from ZFS (noting that $\delta H_{Z}^{L}$ is related to $\delta\phi_Z$ and $\delta A_{\parallel Z}$), while the second term corresponds to contribution from PSZS. 
  
 \subsection{Nonlinear contribution of PSZS to ITG}
 If only PSZS contribution is kept by taking $\delta \phi_{Z}=0$, $\delta A_{\parallel Z}=0$, it results in $\delta H_{Z}^{L}=0$, the corresponding nonlinear particle response to ITG can be derived as   
 \begin{eqnarray}
\delta H_{I,e}^{NL}&=&\frac{e}{T_{e}}F_{Me}\frac{k_{\theta I}}{k_{\theta 0}}\frac{k_{\parallel 0}}{k_{\parallel I}}\delta\hat{\phi}_{0}^{2}\delta\phi_{I},\label{nonlinear electron response to ITG with ZFS}\\
\delta H_{I,i}^{NL}&=&-\frac{e}{T_{i}}\left|\theta_{Z}\right|^{2}F_{Mi}\frac{\omega_{*i}^{t}}{\omega}\delta\hat{\phi}_{0}^{2}\delta\phi_{I}.
\label{nonlinear ion response to ITG with ZFS}
\end{eqnarray}
Substituting equations (\ref{nonlinear electron response to ITG with ZFS}) and (\ref{nonlinear ion response to ITG with ZFS}) into the quasi-neutrality condition, the nonlinear ITG equation in the presence of PSZS beat-driven by TAE can be derived as
\begin{equation}
\begin{aligned} & \left[ \epsilon_{I}-\tau\frac{\omega_{*i}}{\omega}\left(\chi_{iZ}+\eta_{i}\chi_{iZ}-1\right)\delta\hat{\phi}_{0}^{2} \right.\\
&\left.-\frac{k_{\theta I}}{k_{\theta 0}}\frac{k_{\parallel 0}}{k_{\parallel I}}\delta\hat{\phi}_{0}^{2}\right] \delta\phi_{I}\ =\ 0,
\end{aligned}
\label{PSZS on ITG}
\end{equation}
with the term proportional to $\delta \hat{\phi}^2_0$ being the contribution of PSZS beat-driven by TAE.

 \subsection{Nonlinear contribution of ZFS to ITG}
 If we consider only the ZFS contribution by taking $\delta H^{NL}_{Z}=0$, one then obtains the nonlinear particle response to ITG as 
\begin{eqnarray}
\delta H_{I,e}^{NL}&=&-\frac{e}{T_{e}}F_{Me}\frac{k_{\theta I}}{k_{\theta 0}}\frac{k_{\parallel 0}}{k_{\parallel I}}\delta\hat{\phi}_{0}^{2}\delta\phi_{I},\label{nonlinear electron response to ITG with PSZS}\\
\delta H_{I,i}^{NL}&=&\frac{e}{T_{i}}F_{Mi}\left(\left|\theta_{Z}\right|^{2}-1+\frac{\omega_{*i}^{t}}{\omega}\right)\frac{\omega_{*pi}}{\omega}\delta\hat{\phi}_{0}^{2}\delta\phi_{I}.
\label{nonlinear ion response to ITG with PSZS}
\end{eqnarray}
Substituting equations (\ref{nonlinear electron response to ITG with PSZS}) and (\ref{nonlinear ion response to ITG with PSZS}) into the quasi-neutrality condition, the nonlinear ITG equation can be derived as
\begin{equation}
\begin{aligned} & \left[ \epsilon_{I}-\tau\frac{\omega_{*i}}{\omega}\left(-\chi_{iZ}-\eta_{i}\chi_{iZ}+\frac{\omega_{*pi}}{\omega}\right)\delta\hat{\phi}_{0}^{2} \right.\\
&\left.+\frac{k_{\theta I}}{k_{\theta 0}}\frac{k_{\parallel 0}}{k_{\parallel I}}\delta\hat{\phi}_{0}^{2}\right] \delta\phi_{I}\ =\ 0,
\end{aligned}
\label{ZFS on ITG}
\end{equation}
Equation (\ref{ITG WKB D.R.}) can be recovered by combining the nonlinear terms of equations (\ref{PSZS on ITG}) and (\ref{ZFS on ITG}), i.e., the contributions of PSZS and ZFS. We however, will not compare the results of equations (\ref{PSZS on ITG}) and (\ref{ZFS on ITG}) to that of equation (\ref{ITG WKB D.R.}), as the structures of the equations are very different. More specifically, the $1/k_{\parallel I}$ terms in equations (\ref{PSZS on ITG}) and (\ref{ZFS on ITG}) due to non-vanishing nonlinear electron response to ITG, render the corresponding ITG eigenmode equations into a third order differential equation in ballooning space, whereas equation (\ref{ITG WKB D.R.}) will yield a second order differential equation in ballooning space, as the terms proportional to $1/k_{\parallel I}$ in equations (\ref{PSZS on ITG}) and (\ref{ZFS on ITG}) cancel each other. 

\bibliographystyle{aipnum4-1}
\bibliography{ZQiubib}

\end{document}